\documentclass[journal=jacsat,manuscript=article]{achemso}

\usepackage{amsmath}
\usepackage{amssymb}
\usepackage{graphicx}
\usepackage{bm}
\usepackage{dcolumn}

\author{Wangyang Fu}
\author{Cornelia Nef}
\author{Oren Knopfmacher}
\author{Alexey Tarasov}
\author{Markus Weiss}
\author{Michel Calame}
\author{Christian Sch{\"o}nenberger}
  \email{Christian.Schoenenberger@unibas.ch}
\affiliation[University of Basel] {Department of Physics, University of Basel,
Klingel\-berg\-strasse 82, CH-4056 Basel, Switzerland}

\title[\texttt{achemso}]
{Graphene transistors are insensitive to pH changes in solution}

\begin{document}
Version: \date{\today}

\begin{abstract}
We observe very small gate-voltage shifts in the transfer
characteristic of as-prepared graphene field-effect transistors
(GFETs) when the pH of the buffer is changed. This observation is in
strong contrast to Si-based ion-sensitive FETs. The low gate-shift
of a GFET can be further reduced if the graphene surface is covered
with a hydrophobic fluoro\-benzene layer. If a thin Al-oxide layer
is applied instead, the opposite happens. This suggests that clean
graphene does not sense the chemical potential of protons. A GFET
can therefore be used as a reference electrode in an aqueous
electrolyte. Our finding sheds light on the large variety of
pH-induced gate shifts that have been published for GFETs in the
recent literature.
\end{abstract}

Graphene~\cite{Novoselov04} is an extremely interesting and important material with
startling new physical and unique chemical characteristics.\cite{DasSarma2010}
It is a one-atom-thick crystal of carbon atoms arranged into a two-dimensional (2D) hexagonal lattice.
Compared with carbon nano\-tube (CNT), wafer-scale high-quality graphene can readily be
produced by chemical vapor deposition (CVD) on copper and the transfer technique is steadily
improving.\cite{Li09, Li09-2, Bae10,Bhaviripudi10, Lee10} In the straightforward thinking of being used
as a replacement channel material for silicon MOSFET, this unusual simple material has inspired numerous
efforts to fabricate field-effect transistors (FETs).\cite{Avouris07,Chen07,Wang09,Zhang09} Graphene FETs (GFETs)
have also been intensively investigated for high performance chemical and biological
sensors.\cite{Ang08,Ristein10,Cheng10,Ohno09,Heller10,Cohen-Karni10,Chen09,Dankerl10,Levesque11}

The sensing mechanism, however, remains largely unclear, as illustrated
by the large variation of reported pH sensitivities, raging from a low
value of $12$\,mV/pH to a value of $99$\,mV/pH.\cite{Ang08,Ristein10,Cheng10,Ohno09,Heller10}
The latter value is even larger than the thermodynamically allowed maximal shift, the so
called Nernst value ($60$\,mV/pH at room temperature). The present
work serves to resolve this discrepancy.

We demonstrate that the transfer characteristic of as-prepared CVD-grown graphene
shifts surprisingly weakly when the pH of the buffer is changed. The measured value of $6\pm 1$\,mV/pH can
further be reduced to $\sim 0$\,mV/pH when the surface is passivated with a hydrophobic organic layer.
If instead a thin Al-oxide layer is added to graphene, the opposite happens. The pH-induced gate
shift is strongly increased to $17\pm 2$\,mV/pH. This suggests that clean graphene, which in contrast
to oxide surfaces does not expose terminal OH groups to the electrolyte, cannot sense the chemical
potential of protons (the proton concentration), but rather senses the electro\-static potential of the solution.

This finding, that graphene cannot be used to sense the pH of an aqueous solution, is a consequence of its
ideal hydrophobic surface with a very small amount (ideally zero) of dangling bonds. In an aqueous solution
and below the threshold for electrolysis ($\sim 1.25$\,V for water), no specific binding of ions and charge
transfer reaction are expected. In a pioneering work by the McEuen group it was shown that CNT-FETs do not
respond to the chemical, but rather to the electro\-static potential.\cite{Larrimore06}

Our results strongly suggest that the large range of pH-induced gate shifts observed in the previous literature
reflects the quality of graphene. Defective graphene, where free bonds exist on the surface, shows a large shift,
whereas high-quality graphene with no dangling bonds, show no shifts to pH. A clean GFET could therefore act as a
novel solid-state reference electrode that senses only the electrostatic potential in aqueous electrolytes.

%
Figure~1a shows a GFET device ready for measurements in a liquid. GFETs are fabricated in monolayer graphene
grown by CVD on copper and thereafter transferred to a substrate.\cite{Li09,Li09-2,Bae10,Bhaviripudi10,Lee10}
As substrate we use pieces of p-doped Si covered with a $300$\,nm thick thermally grown SiO$_{2}$ layer.
Standard photolithography was applied to define source and drain electrodes made from a bilayer of Ti
($5$\,nm) and Au ($60$\,nm). For the operation in an electrolyte environment, the device was sealed by two additional
fabrication steps.\cite{Knopfmacher10} In the first one, a micrometer-sized liquid channel placed over the graphene is defined
in a photo\-resist layer (AZ2070 nlof) by UV lithography. In the second one, the graphene device was mounted on a chip carrier and wire bonded,
after which an epoxy layer (Epotek 302-3M, Epoxy Technology) was deposited over the contact pads including the bonding
wires. In Figure~1a one can both see the liquid channel (arrow) and the outer epoxy layer (darker
region).

Figure~1b provides a cross-sectional view on the device together with a schematics of the electrical circuit.
As an electrolyte we used standard pH buffer solutions (Tritisol pH 2-9, Merck). The solutions were purged with
pure N$_{2}$ for $2$ hours before the measurements in order to remove any possible influence of O$_{2}$.
As shown schematically in Figure~1b, the gate voltage $V_{Pt}$ was applied to the solution through a Pt wire.
In electro\-chemical terms, the graphene surface is the working electrode, here set to zero potential, and the Pt wire the counter
electrode. The electrostatic potential in solution was monitored by a commercial calomel reference electrode
(REF200, Radiometer analytical) as $V_{ref}$. Electrical conductance measurements were performed at
low source-drain voltage $V_{sd}$ between $10$ and $50$\,mV in the so-called linear regime using a
Keithley 2600A source meter. We have limited the range of applied voltages $V_{Pt}$ to avoid electrolysis.
Moreover, we measured both before and after all pH experiments the electro\-chemical current flowing between the Pt and
the graphene electrodes for the same range of voltages $V_{Pt}$. This current never exceeded $1$\,nA. This is at least
three orders of magnitude smaller than the source-drain current $I_{sd}$ in our experiment. We are therefore confident
that we have measured the conductance change in the graphene layer and not a spurious current.

%
Figure~2 shows the transfer curves of the electrolyte-gated GFET, i.e. the source-drain conductance $G_{sd}$ measured at
$V_{sd}=10$\,mV as a function of the reference gate voltage $V_{ref}$ for different pH buffer solutions. A bipolar transistor
characteristic is observed within an operation voltage of less than $0.8$\,V. This characteristic reflects the fact that the
type of carriers in graphene can continuously be tuned from holes (p-type region to the left in Figure~2) to electrons
(n-type region to the right) by the liquid gate which controls the electro\-chemical potential (Fermi energy)
of the charge carriers. At the transition between the electron and hole regime the conductance is minimal.\cite{Novoselov04}
This point is also referred as the charge-neutrality point (CNP).\cite{Krueger01}
The data shown in the upper part of Figure~2 serves to illustrate the excellent degree of reproducibility. Three subsequent
measurements in the pH=7 buffer, all measured while sweeping $V_{ref}$ from $0.2$ to $0.7$\,V, are shown shifted vertically for
clarity.
For all measured data obtained in different pH buffers a continuous curve is fitted to the data points and used to extract the reference
voltage $V_{CNP}$ for which $G(V_{ref})$ is minimal. The conductance curve $G(V_{ref})$ is nearly symmetric around $V_{CNP}$ and
saturates at large carrier concentration at a value of $350-400$\,$\mu$S, corresponding to a built in series resistance of
$\approx 2.5$\,$\Omega$. The minimum conductance value $G_{min}\approx 250$\,$\mu$S, which corresponds to $6.4$\,$e^2/h$,
was changing slightly with time. Because of this and because of the geometry of the sample, which is not a well defined ribbon, the
value cannot be directly compared with expected minimum conductivities.\cite{Novoselov04}

The dependence of $V_{CNP}$ versus pH is plotted as an inset in Figure~2. The error bars shows the standard deviation deduced from
three consequent measurements. Generally, $V_{CNP}$ shifts with increasing pH (decreasing proton concentration) to higher voltages.
This trend is in agreement with recent reports on GFET pH sensors.\cite{Ang08,Ristein10,Cheng10,Ohno09,Heller10}
Unlike previous work, we find however a considerably reduced value. We extract a sensitivity of $6 \pm 1$\,mV/pH, which is one order of
magnitude smaller than the Nernst limit of $60$\,mV/pH at room temperature.

The observed weak pH sensitivity can be further reduced by adding aromatic benzene-like molecules onto the graphene surface of the
GFET device. It has been shown that the $\pi$-$\pi$ interaction results in a considerable binding energy of almost $0.1$\,eV per
carbon atom for benzene and naphthalene.\cite{Chakarova-Kack06} The idea is schematically sketched in Figure~3. In the experiment we
used fluoro\-benzene molecules. Figure~3 shows the transfer curves of the GFET device measured in different pH buffer solutions after
rinsing the device in flouro\-benzene for $30$\,s and drying. As compared to the as-prepared graphene device, no shift is
discernible. The inset shows $V_{CNP}$ versus pH. Though there is some scatter in the data points, the sensitivity is substantially
reduced to $< 1$\,mV/pH.

Next, we demonstrate the opposite. Up to this point, we have tried to keep the surface as hydrophobic as possible. We now functionalize the
surface with hydroxyl (OH) groups to render it (partially) hydrophilic. This is done by applying a thin Al$_2$O$_3$ layer using
atomic layer deposition (ALD, Savannah 100 from Cambridge NanoTech). We apply a protocol that involves the
activation with NO$_2$ at room temperature before the growth of Al-oxide using Trimethylaluminum (TMA)
as precursor gas.\cite{Farmer06,Williams07} This latter process is an unconventional one, as it was
run at a relatively low temperature of $100$\,$^{\circ}C$. The same process run at the usual temperature of $200$\,$^{\circ}C$
would provide an Al$_2$O$_3$ layer of $\sim 2$\,nm. Due to the non-optimal wetting on graphene
and the substantially reduced growth temperature, the oxide layer is expected to be thinner than $2$\,nm and the coverage may even be
incomplete. The transfer curves for this functionalization looks very interesting. $V_{CNP}$ now shifts quite appreciably when the pH
is changed. The transfer curves are shown in Figure~4 together with $V_{CNP}(pH)$ (inset). A sensitivity of $17$\,mV/pH is deduced. We
emphasize that the curves shift to positive gate voltages for increasing pH values. Such a behavior is expected for an oxide
surface described by the site-binding model.\cite{Bousse83,Bergveld03,vdBerg09} In this model the terminal OH groups on the
surface can be neutral in the form of OH, protonized to OH$_2^{+}$ or de\-protonized to O$^{-}$. At a large pH value (small proton
concentration in solution), the equilibrium is shifted towards a de\-protonized surface which is negatively charged. As a
consequence, the transfer curve is expected to shift to the right. In case of an ideal Al$_2$O$_3$ layer with a large density of
hydroxyl groups a sensitivity of $\sim 40-50$\,mV/pH is expected.\cite{Knopfmacher10} We ascribe the reduced sensitivity
value in our graphene experiment to the low quality of the Al$_2$O$_3$ layer which was grown at a low-temperature. This low
temperature was needed to remain compatible with the organic layers used to seal the devices.


The three experiments measured on one and the same sample, but after different surface treatments, reveal a clear systematics. Adding OH
groups to the surface increases the pH sensitivity. At the same time the transfer curve shifts with increasing pH to larger gate
voltages, a dependence that agrees with the expectation that bound surface charge is the cause for the shifts. If we do the opposite and
apply a hydrophobic coating instead of a hydrophilic one (in our case using fluoro\-benzene), the sensitivity goes to zero. This
suggests that the fluro\-benzene molecules are able to suppress the chemical activity of residual free bonds on the as-grown graphene
surface which by itself was showing a pH response, albeit a small one. This picture immediately leads to the conclusion that the ideal
defect free graphene surface should yield no response to a change of the pH. This is expected because the surface can neither be
protonized nor de\-protonized. Hence, it cannot sense the proton concentration in solution. Any sign of pH response must be taken as
a sign of imperfection.

In the previous literature there was a contradiction in the observed sensitivity of graphene transistors operating as pH sensors. Ang
\textit{et al.} reported a super-Nernstian response of $99$\,mV/pH in single-, double- and triple-layer graphene FETs.\cite{Ang08}
However, this superior value could be simply due to the large gate leakage current of their GFETs, as commented recently
by Ristein \textit{et al}.\cite{Ristein10} Cheng \textit{et al.} reported a pH sensitivity of $\sim 20$\,mV/pH in suspended graphene transistor
with reduced noise.\cite{Cheng10} However, the unsealed device design indicates the direct exposure of bare Au electrodes to
liquid, which raises uncertainty in their analysis also because of possible leakage currents. Heller \textit{et al.} recently presented
a comprehensive study on the influence of the electrolyte composition on liquid-gated CNT and graphene transistors.\cite{Heller10}
They observed a comparably weak pH sensitivity of $\sim 9$\,mV/pH and $\sim 12$\,mV/pH for CNT and graphene in $1$\,M KCl solution,
respectively. However, if the concentration of KCl is reduced, a comparably large sensitivity close to $50$\,mV/pH has been reported.
Still, in several other recent publications, values around $20$\,mV/pH were reported for graphene.\cite{Ohno09}

Our results convincingly demonstrate that the weak pH sensitivity is not a surprise for graphene whose surface has saturated carbon
bonds. Therefore, no specific binding of ions is expected in the ideal case of a clean graphene surface. In practice, however, there
will always be some defects on the surface and along the edges, which are induced during transfer and device fabrication. These
defects, hydroxyl and carbonyl groups, for example\cite{Mkhoyan09}, can react with the protons in the electrolyte, yielding a spurious
pH signal.

In summary, we have systematically studied the response of electrolyte-gated graphene FETs to different pH solutions and
demonstrated that graphene is intrinsically inert to pH. As-fabricated graphene FETs have a weak pH sensitivity of
\mbox{$6$\,mV/pH}. This weak pH sensitivity of graphene can be further reduced by passivating the surface with hydrophobic
fluoro\-benzene molecules. An appreciable pH response can be induced if the surface is covered with a thin inorganic oxide layer. This
layer provides terminal hydroxyl groups which can be protonized and de\-protonized yielding a bound surface charge layer whose charge
density depends on the proton concentration in solution. Our results suggest that the observed relatively large pH sensitivities reported
in the literature for graphene can be understood by considering graphene of different quality. Ideal defect-free graphene should not
respond to pH, whereas defective one will. Graphene FETs are therefore not suited for pH sensors. In contrast, graphene is the
ideal platform for reference electrodes to probe the electrostatic potential in an aqueous electrolyte. Since reference electrodes are
typically bulky objects, they are not easily integrated. For dynamical binding and unbinding experiments in liquid channels,
integrated reference electrodes are crucial elements. Graphene transistors hold therefore great promise in the use as a novel
integrateable solid-state reference electrodes.

\begin{acknowledgement}
  The authors acknowledge funding from the Swiss Nanoscience Institute (SNI), nanotera.ch,
  Swiss NSF and the ESF programme Euro\-graphene.
\end{acknowledgement}

%
%



\newpage

\begin{thebibliography}{26}


\bibitem{Novoselov04}
  Novoselov, K. S.; Geim, A. K.; Morozov, S. V.; Jiang, D.; Zhang, Y.; Dubonos, S. V.; Grigorieva, I. V.; Girsov, A. A.
  \textit{Science} \textbf{2004}, \textsl{306}, 666-669.

\bibitem{DasSarma2010}
  Das Sarma, S.; Adam, S.; Hwang, E. H.; Rossi, E.
  \emph{arXiv:1003.4731v2} (2010).

\bibitem{Li09}
  Li, X. S.; Cai, W. W.; An, J.; Kim, S.; Nah, J.; Yang, D. X.; Piner, R.; Velamakanni, A.;
  Jung, I.; Tutuc, E.; Banerjee, S. K.; Colombo, L.; Ruoff, R. S.
  \textit{Science} \textbf{2009}, \textsl{324}, 1312-1314.

\bibitem{Li09-2}
  Li, X. S.; Zhu, Y. W.; Cai, W. W.; Borysiak, M.; Han, B.; Chen, D.; Piner, R. D.; Colombo, L.; Ruoff, R. S.
  \textit{Nano Lett.} \textbf{2009}, \textsl{9}, 4359-4363.

\bibitem{Bae10}
  Bae, S.; Kim, H.; Lee, Y.; Xu, X. F.; Park, J. S.; Zheng, Y.; Balakrishnan, J.; Lei, T.; Kim, H. R.; Song, Y. I.; Kim, Y. J.;
  Kim, K. S.; {\"O}zyilmaz, B.; Ahn, J. H.; Hong, B. H.; Iijima, S.
  \textit{Nature Nanotech.} \textbf{2010}, \textsl{5}, 574-578.

\bibitem{Bhaviripudi10}
  Bhaviripudi, S.; Jia, X. T.; Dresselhaus, M. S.; Kong, J.
  \textit{Nano Lett.} \textbf{2010}, \textsl{10}, 4128-4133.

\bibitem{Lee10}
  Lee, S.; Lee, K.; Zhong, Z. H.
  \textit{Nano Lett.} \textbf{2010}, \textsl{10}, 4702-4707.

\bibitem{Avouris07}
  Avouris, Ph.; Chen, Z. H.; Perebeinos, V.
  \textit{Nature Nanotech}. \textbf{2007}, \textsl{2}, 605-615.

\bibitem{Chen07}
  Chen, Z. H.; Lin, Y. M.; Rooks, M. J.; Avouris, Ph.
  \textit{Physica E} \textbf{2007}, \textsl{40}, 228-232.

\bibitem{Wang09}
  Wang, X. R.; Li, X. L.; Zhang, L.; Yoon, Y.; Weber, P. K.; Wang, H. L.; Guo, J.; Dai, H. J.
  \textit{Science} \textbf{2009}, \textsl{324}, 768-771.

\bibitem{Zhang09}
  Zhang, Y. B.; Tang, T. T.; Girit, C.; Hao, Z.; Martin, M. C.; Zettl, A.; Crommie, M. F.; Shen, Y. R.; Wang, F.
  \textit{Nature} \textbf{2009}, \textsl{459}, 820-823.


\bibitem{Ang08}
  Ang, P. K.; Chen, W.; Wee, A. T. S.; Loh, K. P.
%
  \textit{J. Am. Chem. Soc.} \textbf{2008}, \textsl{130}, 14392-14393.

\bibitem{Ristein10}
  Ristein, J.; Zhang, W. Y.; Speck, F.; Ostler, M.; Ley, L.; Seyller, T.
  \textit{J. Phys. D: Appl. Phys.} \textbf{2010}, \textsl{43}, 345303.

\bibitem{Cheng10}
  Cheng, Z. G.; Li, Q.; Li, Z. J.; Zhou, Q. Y.; Fang, Y.
  \textit{Nano Lett.} \textbf{2010}, \textsl{10}, 1864-1868.

\bibitem{Ohno09}
  Ohno, Y.; Maehashi, K.; Yamashiro, Y.; Matsumoto, K.
  \textit{Nano Lett.} \textbf{2009}, \textsl{9}, 3318-3322.

\bibitem{Heller10}
  Heller, I.; Chatoor, S.; M{\"a}nnik, J.; Zevenbergen, M. A. G.; Dekker, C.; Lemay, S. G.
  \textit{J. Am. Chem. Soc.} \textbf{2010}, \textsl{132}, 17149-17156.

\bibitem{Cohen-Karni10}
  Cohen-Karni, T.; Qing, Q.; Li, Q.; Fang, Y.; Lieber, C. M.
  \textit{Nano Lett.} \textbf{2010}, \textsl{10}, 1098-1102.


\bibitem{Chen09}
  Chen, F.; Xia, J. L.; Tao, N. J.
  \textit{Nano Lett.} \textbf{2009}, \textsl{9}, 1621-1625.

\bibitem{Dankerl10}
  Dankerl, M.; Hauf, M. V.; Lippert, A.; Hess, L. H.; Birner, S.; Sharp, I. D.; Mahmood, A.; Mallet, P.; Veuillen, J. Y.; Stutzmann, M.; Garrido, J. A.
  \textit{Adv. Funct. Mater.} \textbf{2010}, \textsl{20}, 3117-3124.

\bibitem{Levesque11}
  Levesque, P. L.; Sabri, S. S.; Aguirre, C. M.; Guillemette, J.; Siaj, M.; Desjardins, P.; Szkopek, T.; Martel, R.
  \textit{Nano Lett.} \textbf{2011}, \textsl{11}, 132-137.

\bibitem{Larrimore06}
  Larrimore, L.; Nad, S.; Zhou, X. J.; Abru$\tilde{n}$a, H.; McEuen, P. L.
  \textit{Nano Lett.} \textbf{2006}, \textsl{6}, 1329-1333.


\bibitem{Knopfmacher10}
  Knopfmacher, O.; Tarasov, A.; Fu, W. Y.; Wipf, M.; Niesen, B.; Calame, M.; Sch{\"o}nenberger, C.
  \textit{Nano Lett.} \textbf{2010}, \textsl{10}, 2268-2274.

\bibitem{Krueger01}
  Kr{\"u}ger, M.; Buitelaar, M. R.; Nussbaumer, T.; Sch{\"o}nenberger, C.; Forr{\'o}, L.
  \textit{Appl. Phys. Lett.} \textbf{2001}, \textsl{78}, 1291-1293.

\bibitem{Chakarova-Kack06}
  Chakarova-K{\"a}ck, S. D.; Schr{\"o}der, E.; Lundqvist, B. I.; Langreth, D. C.
  \textit{Phys. Rev. Lett.} \textbf{2006}, \textsl{96}, 146107.

\bibitem{Farmer06}
  Farmer, D. B.; Gordon, R. G.
  \textit{Nano Lett.} \textbf{2006}, \textsl{6}, 699-703.

\bibitem{Williams07}
  Williams, J. R.; DiCarlo, L.; Marcus C. M.
  \textit{Science} \textbf{2007}, \textsl{317}, 638-641.

\bibitem{Bousse83}
  Bousse, L.; De Rooij, N. F.; Bergveld, P.
  \textit{IEEE Trans. Electron Dev.} \textbf{1983}, \textsl{ED-30}, 1263-1270.

\bibitem{Bergveld03}
  Bergveld, P.
  \textit{Sens. Actuators B} \textbf{2003}, \textsl{88}, 1-20.

\bibitem{vdBerg09}
  Chen, S. Y.; Bomer, J. G.; van der Wiel, W. G.; Carlen, E. T.; van den Berg, A.
  \textit{ACS Nano} \textbf{2009}, \textsl{3}, 3485-3492.

\bibitem{Mkhoyan09}
  Mkhoyan, K. A.; Contryman, A. W.; Silcox, J.; Stewart, D. A.; Eda, G.; Mattevi, C.; Miller, S.; Chhowalla, M.
  \textit{Nano Lett.} \textbf{2009}, \textsl{9}, 1058-1063.


\end{thebibliography}

\newpage

\begin{figure}[htb]
\begin{center}
\includegraphics[width=160mm]{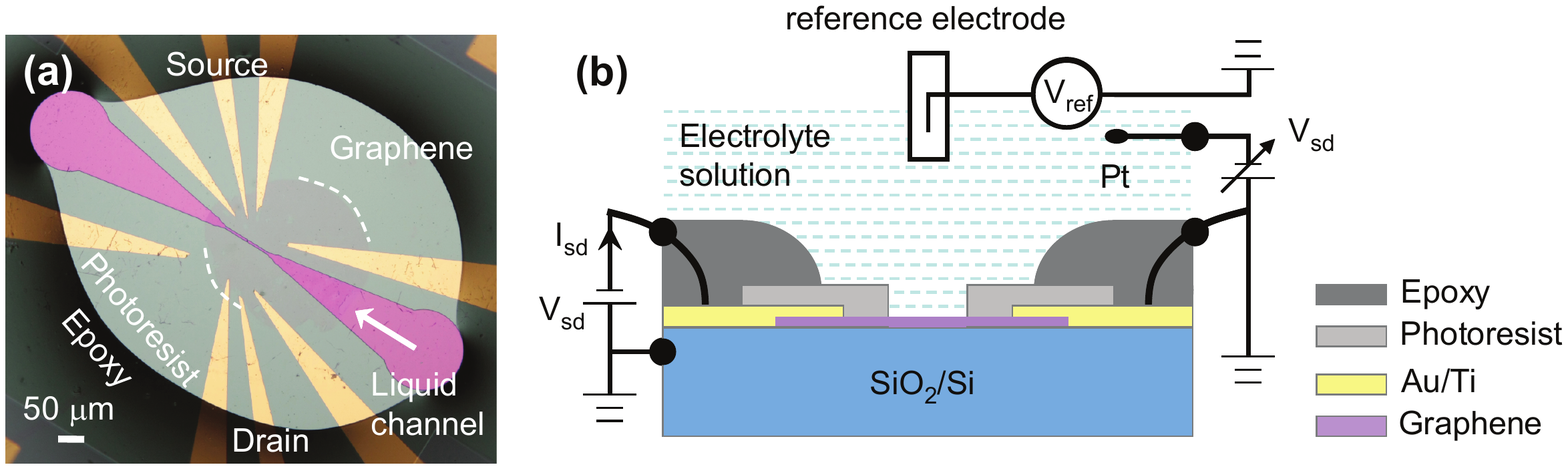}
\end{center}
\caption{(a) An optical image of large-sized graphene transistor (GFET)
with Ti/Au source and drain contacts. The device, including the
bonds, is sealed except for a liquid channel (arrow) running across the
graphene sheet, whose boarder is shown by dashed curves.
(b) Schematics of the experimental setup and the electrical circuitry of
the electrolyte-gated GFET. The gate voltage $V_{Pt}$ was applied to
the solution via a Pt wire. The electrostatic potential in solution  $V_{ref}$
was monitored by the reference calomel electrode}
  \label{fig1}
\end{figure}

\newpage

\begin{figure}[htb]
\begin{center}
\includegraphics[width=120mm]{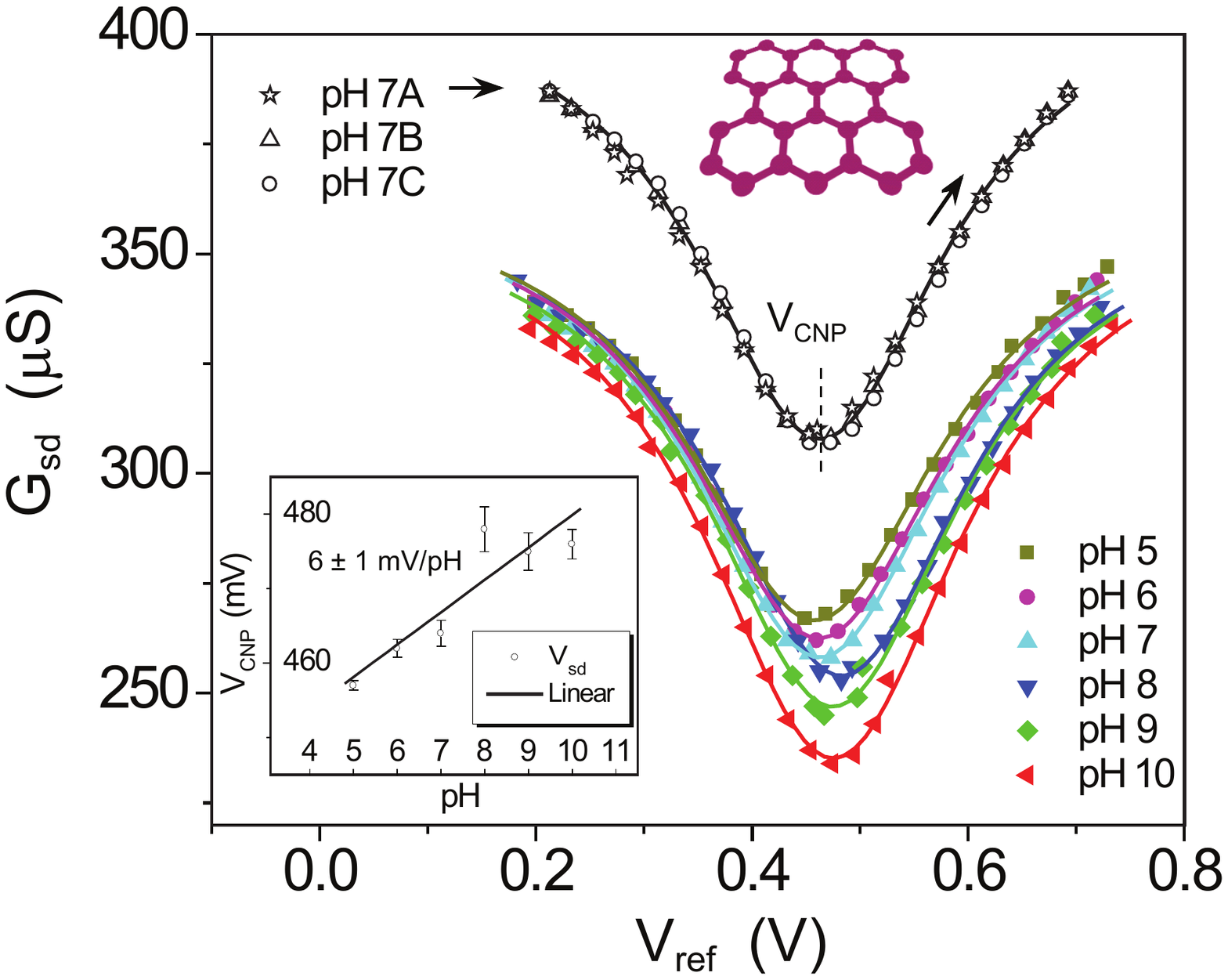}
\end{center}
  \caption{Electrical source-drain conductance $G_{sd}$ as a function of
  the reference potential $V_{ref}$ measured in different pH buffer solutions for the
  as-prepared GFET. A bipolar transfer curve is observed corresponding to
  different type of charge carriers that can continuously be tuned from holes (left) to electrons (right)
  with the charge-neutral point V$_{CNP}$ at minimum $G_{sd}$.
  The transfer curve shift slightly to more positive $V_{ref}$ with increasing pH.
  Inset: A sensitivity of $6\pm 1$\,mV/pH is deduced. Error bars represent the
  standard deviation from three subsequent measurements taken for each pH value.
  As an example, the three data sets obtained for pH=7 are explicitly shown on the top (vertically shifted for clarity)}
 \label{fig2}
\end{figure}

\newpage

\begin{figure}[htb]
\begin{center}
\includegraphics[width=120mm]{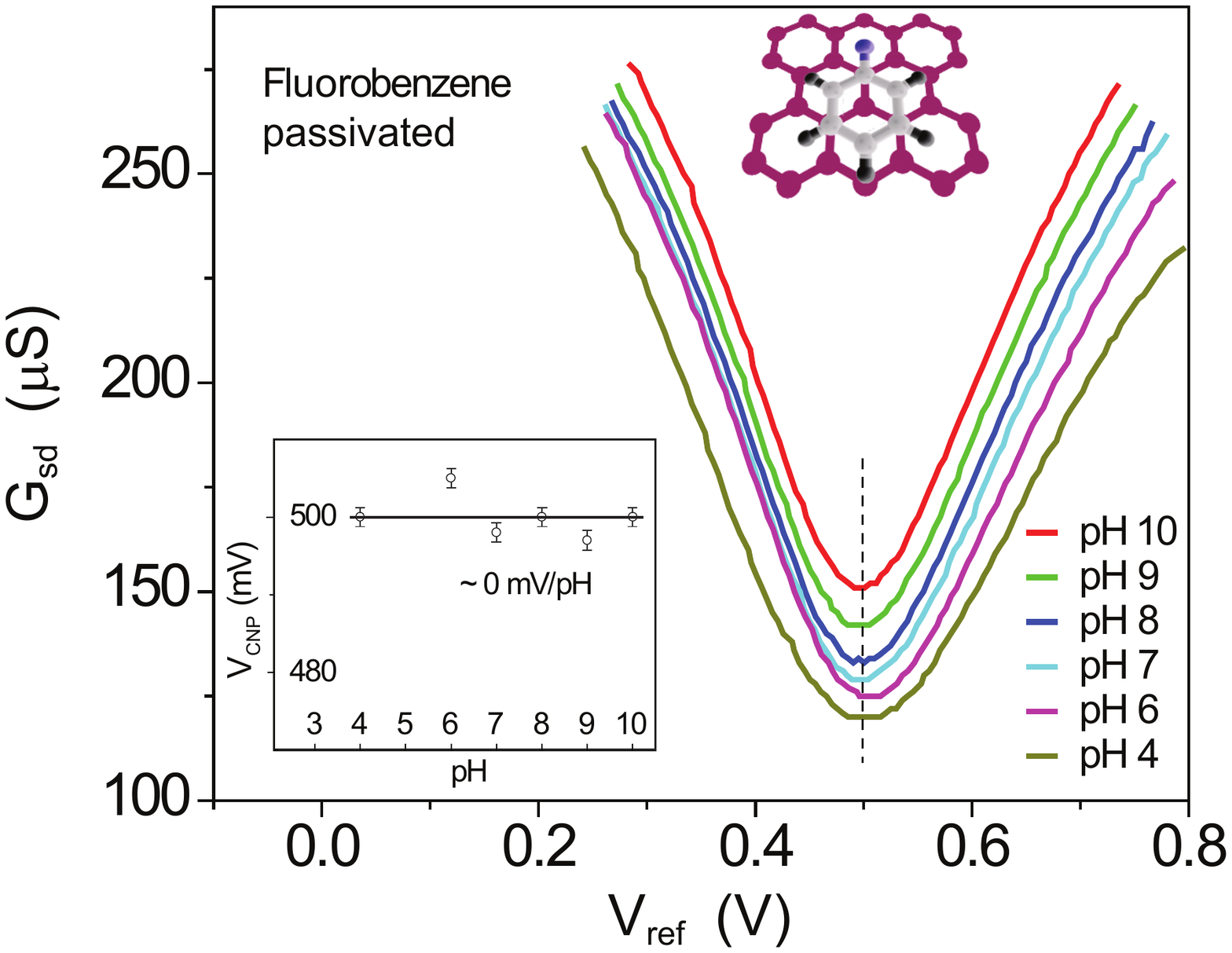}
\end{center}
  \caption{Electrical source-drain conductance $G_{sd}$ as a function of
  the reference potential $V_{ref}$ measured in different pH buffer solutions for
  the GFET device after rinsing in fluoro\-benzene for $30$\,s and drying.
  Inset: The transfer curves do not shift at all when changing the pH.}
  \label{fig3}
\end{figure}

\newpage

\begin{figure}[htb]
\begin{center}
\includegraphics[width=120mm]{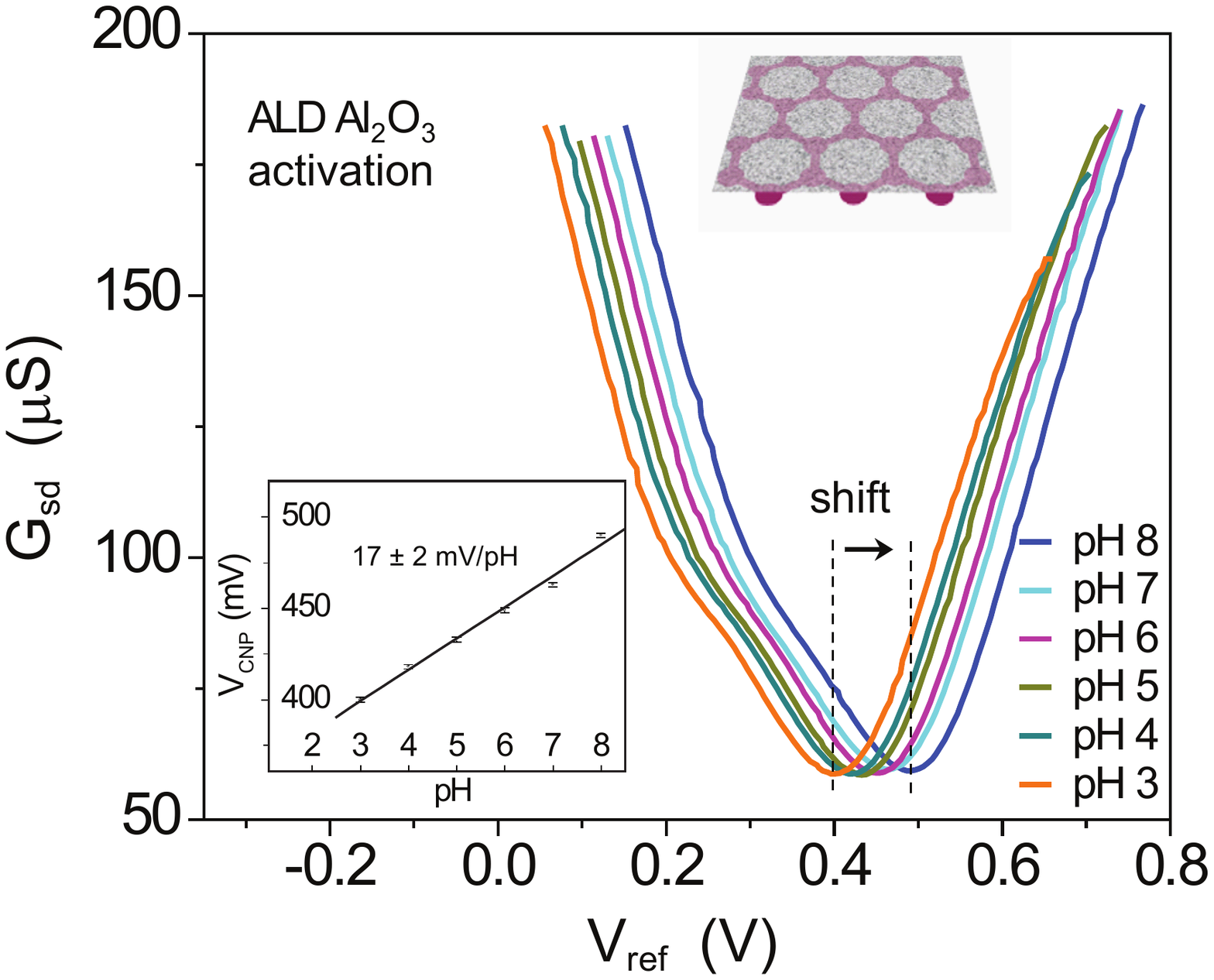}
\end{center}
  \caption{Electrical source-drain conductance $G_{sd}$ as a function of
  the reference potential $V_{ref}$ measured in different pH buffer solutions for
  the GFET device after having applied a thin Al$_{2}$O$_{3}$ coating by ALD.
  There is now an appreciable shift of the transfer curves to more positive $V_{ref}$ with increasing pH.
  Inset: A sensitivity of $17 \pm 2$\,mV/pH is deduced.}
  \label{fig4}
\end{figure}

\end{document}